\documentclass{article}
\usepackage{authblk}
\usepackage[english]{babel}
\usepackage[utf8]{inputenc}
\usepackage{graphicx,epstopdf}
\usepackage[final]{pdfpages}
\usepackage{cool}
\usepackage{amssymb}
\usepackage{amsmath}
\usepackage{amsthm}
\usepackage{amsfonts}
\usepackage{amscd}
\usepackage[breaklinks]{hyperref}
\usepackage{url}
\usepackage{breakurl}
\usepackage[margin=1.1in]{geometry}

\usepackage{natbib}

\allowdisplaybreaks

\title{On new data sources for the production of official statistics}
\author[1,2]{David Salgado}
\author[3]{Bogdan Oancea}
\affil[1]{Dept.\ Methodology and Development of Statistical Production, Statistics Spain (INE), Spain}
\affil[2]{Dept.\ Statistics and Operations Research, Complutense University of Madrid, Spain}
\affil[3]{Dept.\ Business Administration, University of Bucharest, Romania}
\date{February 7, 2020}

\begin{document}





\maketitle

\begin{abstract}
In the past years we have witnessed the rise of new data sources for the potential production of official statistics, which, by and large, can be classified as survey, administrative, and digital data. Apart from the differences in their generation and collection, we claim that their lack of statistical metadata, their economic value, and their lack of ownership by data holders pose several entangled challenges lurking the incorporation of new data into the routinely production of official statistics. We argue that every challenge must be duly overcome in the international community to bring new statistical products based on these sources. These challenges can be naturally classified into different entangled issues regarding access to data, statistical methodology, quality, information technologies, and management. We identify the most relevant to be necessarily tackled before new data sources can be definitively considered fully incorporated into the production of official statistics.
\end{abstract}

\tableofcontents

\section{Introduction}
In October 2018, the 104th DGINS conference \citep{DGINS18a}, gathering all directors general  of the European Statistical System (ESS), \textquotedblleft [a]gree[d] that the variety of new data sources, computational paradigms and tools will require amendments to the statistical business architecture, processes, production models, IT infrastructures, methodological and quality frameworks, and the corresponding governance structures, and therefore invite[d] the ESS to formally outline and assess such amendments\textquotedblright. Certainly, this statement is valid for producing official statistics in any statistical office.\\

More often than not, this need for the modernisation of the production of official statistics is associated with the rising of \emph{Big Data} \citep[e.g.][]{DGINS13a}. In our view, however, this need is also naturally linked to the use of administrative data \citep[e.g.][]{ESS13a} and even earlier to the efforts to boost the consolidation of an international industry for the production of official statistics through shared tools, common methods, approved standards, compatible metadata, joint production models and congruent architectures \citep{HLGMOS11a, UNE19a}.\\

Diverse analyses can be found in the literature providing insights about the challenges of Big Data and new digital data sources, in general, for the production of official statistics \citep{StrBraDaa14a, Lan14a, AAPOR15a, ReiKou15a, Kit15a, IMF17a, GicSzo18a, BraZee20a}. These analyses are mostly strategic, high-level, and top-down. In this work we undertake a bottom-up approach mainly aiming at identifying those factors underpinning the reason why statistical offices are not producing outputs based on all these new data sources yet. Simply put: why are statistical offices not producing routinely official statistics based on these new digital data sources?\\

Our main thesis is that for statistical products based on new data sources to become routinely disseminated according to updated legal national and international regulations, at least, all the issues identified below must be provided with a widely acceptable solution. Should we fail to cope with the challenges behind one of these issues, the new products cannot be achieved. Thus, we are facing an intrinsically multifaceted problem. Furthermore, we shall argue that new data sources are compelling a new role of statistical offices derived from the social, statistical, and technical complexity of the new challenges.\\

These challenging issues are discussed separately in each section. In section \ref{sec:Data} we revise relevant aspects of the concept of data and its implications to produce official statistics. In section \ref{sec:Access} we tackle the issue of access to these new data sources. In section \ref{sec:Method} we briefly identify issues regarding the new statistical methodology necessary to undertake the production with both the traditional and the new data. In section \ref{sec:Qual} we deal with the implications regarding the quality assurance framework. In section \ref{sec:IT} we shortly approach the questions about the information technologies. In section \ref{sec:SkillHR} we pose some reflections regarding skills, human resources, and management in statistical offices. We close with some conclusions in section \ref{sec:Con}.

\section{Data: survey, administrative, digital}
\label{sec:Data}

\subsection{Some definitions}
\label{sub:Definitions}
The production of official statistics is a multifaceted concept. Many of these facets are affected by the nature of the data. We pose some reflections about some of them. In a statistical office three basic data sources are nowadays identified: survey, administrative, digital. This distinction runs parallel to the  historical development of data sources.\\

A survey is \textquotedblleft scientific study of an existing population of units typified by persons, institutions, or physical objects\textquotedblright\ \citep{LesKal92a}. This is not to be confused with the idea of sampling itself, introduced in Official Statistics in 1895 by A.\ Ki{\ae}r as the \emph{representative method} \citep{Kia87a}, provided with a solid mathematical basis and promoted to probability sampling firstly by \citet{Bow06a} and definitively by \citet{Ney34a}, further developed originally in the US Census Bureau \citep{Dem50a,HanHurMad66a, Han87a} \citep[see also ][]{Smi76a}, and still in common practice by statistical offices worldwide \citep{Bet09a,Bre13a}. It has been the preferred and traditional tool to elaborate and produce official information about any finite population. The advent of different technologies in the 20th century produced a proliferation of so-called data collection modes (CAPI, CATI, CAWI, EDI, etc.) \citep[cf. e.g.][]{BieLyb03a}, but the essence of a survey is still there.\\

Administrative data is \textquotedblleft the set of units and data derived from an administrative source\textquotedblright, i.e.\ from an \textquotedblleft organisational unit responsible for implementing an administrative regulation (or group of regulations), for which the corresponding register of units and the transactions are viewed as a source of statistical data\textquotedblright\  \citep{OECD08a}. Some experts \citep[see Deliverable 1.3 of][]{ESS13a} drop the notion of units to avoid potential confusion and just refer to data. All in all, these definitions refer to registers developed and maintained for administrative and not statistical purposes. Apart from the diverse traditions in countries for the use of these data in the production of official statistics, in the European context the Regulation No.\ 223/2009 provides the explicit legal support for the access to this data source by national statistical offices for the development, production and dissemination of European statistics \citep[see Art.\ 24 of][]{EC09a}. Curiously enough, the Kish tablet from the Sumerian empire (ca 3500 bC), one of the earliest examples of human writing, seems to be an administrative record for statistical purposes.\\

More recently, the proliferation of digital data in an increasing number of human activities has posed the natural challenge for statistical offices to use this information for the production of official statistics. The term \emph{Big Data} has polarised this debate with the apparent abuse of the $n$ Vs definitions \citep{Lan01a, Nor13a}. But the phenomenon goes beyond this characterization extending the potentiality for statistical purposes to any sort of digital data. In parallel to administrative data, we propose to define digital data as \emph{the set of units and data derived from a digital source}, i.e.\ from  a \emph{digital information system, for which the associated databases are viewed as a source of statistical data}.\\

Notice that the \citet{OECD08a} does not include this as one of the types of data sources, probably because this definition of digital data may be read as falling within the more general one of administrative data above, since administrative registers are nowadays also digitalised. We shall agree on restricting administrative data to the public domain, in agreement with current practice in statistical offices and in application of EU Regulation 223/09.\\

This discrimination between the three sources of data runs parallel to their collection modalities: surveys are essentially collected through structured interviews administered directly to the statistical unit of interest, administrative registers are collected from public administrative units, and digital data offers an undefined variety of potential private data providers (either individual or organizational). However, we want to emphasise that differences among these data sources are deeper than just their collection modalities. Furthermore, these differences lie at the core of many of the challenges described in the next sections.

\subsection{Statistical metadata}
\label{sub:StatMeta}
The first determining factor for the differences in these data sources is the presence/absence of \textbf{statistical metadata}, i.e.\ metadata for statistical purposes. Not only is it relevant to understand what this means but also especially to identify the reason why this is introducing differences. Data such as survey data generated with statistical structural metadata embrace variables following strict definitions directly related to target indicators and aggregates under analysis (unemployment rates, price indices, tourism statistics, \dots). These definitions are operationalised in careful designs of questionnaire items. Data are processed using survey methodology, which provides a rigorous inferential framework connecting data sets with target populations at stake.\\

On the contrary, data such as administrative and digital data generated without statistical structural metadata embrace variables with a faint connection with target indicators and aggregates. This impinges on their further processing in many aspects and especially regarding data quality and the inference with respect to target populations. The ultimate reason for this absence of statistical metadata is that this data is generated to provide a non-statistical transactional service (taxes, medical care benefits, financial transactions, telecommunication, \dots). This has already been identified in the literature \citep{Han18a}. In contraposition to survey data, administrative and digital data are generated before their corresponding statistical metadata. They do have metadata, but not for statistical purposes.\\

In our view, the key distinguishing factor, derived from this absence of statistical metadata, arises from the explicit or (mostly) implicit conception of information behind data. This plays a critical role in the statistical production process. The concept of information gathers three complementary aspects, namely (i) syntactic aspects concerning the quantification of information, (ii) semantic problems related to meaning, and (iii) utility issues regarding the value of information \citep[see e.g.][]{Flo19a}. When considering the traditional production of official statistics, we all are aware of the substantial investment on the system of metadata providing rigorous and unambiguous definitions for each of the variables collected in a survey, work conducted \textbf{prior to data collection}. This is providing survey data with a purposive semantic layer and noticeably increasing its value (all three aspects of the concept of information meet in survey data). On the contrary, administrative data are not generated under this umbrella of statistical metadata, but their semantic content is often still close enough to the statistical definitions used in a statistical office (think e.g.\ of the notions of employment, taxes, education grades, etc.). Nonetheless, the quality of administrative data for statistical purposes is still an issue \citep[see e.g.][]{AgaGraKloReiVaj15a,FolShuMar18a,KelMulMorKon18a}. The situation with digital data is extreme. This data is generated to provide some kind of service completely extraneous to statistical production. Thus, meaning and value must be carefully worked out for the new data to be used in the production of official statistics (only the first layer of the concept of information is present in digital data). Some proposed architectures for the incorporation of new data sources \citep{Ric18a, Eur19a} reflect this situation: a non-negligible amount of preprocessing is required prior to incorporate digital data into the statistical production process.\\

This different informational content of data for producing official statistics will prove to have far-reaching consequences on the production methodology. We can borrow a well-known episode in the History of Science to illustrate this difference and its consequences: the Copernican scientific revolution substituting the Ptolemaic system by the Newtonian law of universal gravitation \citep{Kuh57a}. The Ptolemaic system enables us to compute and predict the behaviour of any celestial body by introducing more and more computational elements such as epicycles and deferents. Newton's law of universal gravitation also enables us to compute this behavior under a completely different perspective. We can assimilate the former with a purely syntactic usage of data whereas the latter is somewhat incorporating meaning (theory). This is not a black-or-white comparison, since there is some theory behind the Ptolemaic system (Aristotelian Physics) but the difference in the comprehension of natural phenomena provided by both systems is appealing, even using the same set of data. In other words, in the former case we just introduce our observed astronomical data into a more or less entangled computation system whereas in the latter case we make use of underlying assumptions providing context, meaning, and explanations for all the observed data. In an analogous way, let us now consider the difference between a regression model and a random forest, also for the same set of (big) data. In the former, some meaning is incorporated or at least postulated through the choice of a functional form between regressand and regressors (linear, logistic, multinomial, etc.). In the latter, only weaker computational assumptions are made. The situation is similar to the cosmological picture above and indeed lies at the dichotomy between the so-called \textquotedblleft theory-driven\textquotedblright\ and \textquotedblleft data-driven\textquotedblright\ approaches to data analysis \citep[see e.g.][]{Han19a}. This also runs parallel to the model-based vs.\ design-based inference \citep{Smi94a}, whose finally adopted solution in favour of the latter can be summarised with the following statement by \citet[][p. 785]{HanMadTep83a}: \textquotedblleft [\dots] it seems desirable, to the extent feasible, to avoid estimates or inferences that need to be defended as judgments of the analysts conducting the survey\textquotedblright. Avoiding prior hypotheses about data generation is possible using probability sampling (survey data), but with new data sources this is not the case anymore. This duality has already been identified in the use of Big Data as the historical debate between rationalism and empiricism \citep{Sta16a}.\\

Thus, as a challenging issue, we may enquire whether statistical offices should still adopt a merely computational (empiricist) point of view \emph{\`{a} la} Ptolomy or should they pursue theoretical (rationalistic) findings \emph{\`a la} Newton perhaps searching for a better system of computation and estimation. No clear position is recognised in our community yet and this will impinge not only on the statistical methodology for new data sources but on the whole role of statistical offices in society. This change will indeed be very deep.

\subsection{Economic value}
\label{sub:EcoVal}

The second main difference among the three data sources arises from their economic value. Traditional survey data has little economic value for a data holder/provider in comparison with digital data. For example, when a company owing a database for an online job vacancy advertisement service is requested to provide data about their turnover, number of employees, R+D investment, etc., sharing this information does not reasonably seem to be as critical as sharing this whole database for official statistical production. In the case of administrative data, whose public dimension we agreed upon above, the economic value for the public administration is secondary (statistical offices are indeed part of the public administration).\\

This economic value entails diverse consequences for the incorporation of digital data sources into official statistics production. Data collection is clearly more demanding. On the one hand, technological challenges lie ahead about retrieving, preprocessing, storing, and/or transmitting these new databases. On the other hand, and more importantly, by accessing to the business core of data holders, the degree of disruption of official statistical production into their business processes is certainly higher. Moreover, technical staff is usually required to access these data sources and even to preprocess and interpret them for statistical purposes (e.g. telco data). This also impinges directly on the capability profiles of official statisticians. Thus, it is appealingly different to collect (either paper or electronic) questionnaires than to access huge business databases.\\

The economic value of digital data constitutes a key feature which demands careful attention by national and international statistical systems. The perception of risk for e.g.\ settling public-private partnerships \citep{RobKleJut15a} runs indeed parallel to this economic value. High economic value comes usually as a result of high investments, therefore sharing core business data with statistical offices may be easily perceived as too high a risk. However, if these public-private partnerships are perceived as an opportunity to increase this economic value (increasing e.g. data quality, the quality of commercial statistical products, and the social dimension of private economic activities), the statistical production and the information and knowledge generation thereof can be reinforced in society.\\

As we shall discuss in a later section, this suggests to broaden the scope of official statistical outputs from traditionally closed and embedded in statistical domains (usually according to a strict legal regulation) to some enriched intermediate high-quality datasets for further customised production by other economic and social actors (researchers, companies, NGOs, \dots) in a variety of socioeconomic domains. This is also a deep change in National Statistical Systems.

\subsection{Third people}
\label{sub:ThiPeo}
The third main difference stems out from the fact that these digital data refer to third people, not to data holders themselves. These third people are clients, subscribers, etc.\ sharing their private information in return of a business service. Implications immediately arise. Issues about the legal support for access are obvious (see section \ref{sec:Access}), but this factor is not entirely new. In survey methodology we already have the notion of proxy respondent \citep[see e.g.][]{Cob18a} and in administrative data, information about citizens and not about the data-holding public institutions is the core of this data source.\\

Confidentiality and privacy issues naturally arise. Already in the traditional official statistical process a whole production step is dedicated to statistical disclosure control \citep{HunDomFerFraGieNorSpideW12a} reducing the re-identification risk of any sampling unit while assuring the utility of the disseminated statistics. Now, the data deluge has made this risk increase since it is more feasible to identify individual population units \citep{RocHendeM19a}, even despite data are not personally identified anymore (in contrast to survey and administrative data). Apart from the spread of privacy-by-design statistical processes, now more advanced cryptographical techniques such as Multiparty Secure Computation \citep[see][and multiples references therein]{ZhaZhaZhaCheGaoLiTan19a} must be taken into consideration, especially regarding data integration.\\

Ethical issues should also be considered. Long can be written about ethics of requesting private information to both people or enterprises in a general setting. Regarding new data sources, the debate about accessing data for the production of official statistics has received attention, in particular regarding privacy and confidentiality. Since we do not have a clear position regarding this issue, we just want to provocatively share two reflections. Firstly, with both survey and administrative sources, data for official statistics is personal data where either people and business units are univocally identified in internal sets of microdata at statistical offices. Take for example the European Health Survey \citep{EHS}. Items like the following are included in the questionnaires:

\begin{itemize}
\item[\textbf{HC08}] When was the last time you visited a dentist or orthodontist on your own behalf (that is, not
while only accompanying a child, spouse, etc.)?
\item[\textbf{HC09}] During the past four weeks ending yesterday, that is since (date), how many times did you
visit a dentist or orthodontist on your own behalf?
\end{itemize}

This sensitive information is collected together with a full identification of each respondent. Other example is a historical and fundamental statistics for society: causes of death \citep{Eur20a}. People committing suicide or dead due to alcoholic abuse, for example, are clearly identified in internal sets of microdata at statistical offices. Most digital sources provide anonymous (or pseudo-anonymous) data. Notice that for the case of the European Health Survey even duly anonymised microdata are publicly shared. Have statistical offices not been careful enough so far in the application of IT security and statistical disclosure control to scrupulously protect both privacy and confidentiality of statistical units in their traditional statistical products? Uses other than strictly statistical purposes have not been followed in the usage of this information by statistical offices. Even despite the risk of identifiability, should the production of official statistics revise the ethics of its activity? Even for traditional sources?\\

Secondly, the fast generation of digital data nowadays clearly poses an immediate question. Should or should not society elaborate accurate and timely information for those matters of public interest (CPI, GDP, unemployment rates, \dots and even potentially novel insights) taking advantage of this data deluge? In all cases this is posed in full compatibility with increasing economic sectors around digital data both for statistical and nonstatistical purposes.\\

Obviously, this debate is part of the social challenges behind the generation of such an amount of digital information. Statistical offices cannot be aside and should assume their role.

\section{Access}
\label{sec:Access}
Needless to say, should statistical offices have no access to new digital data sources, no official statistical product can be offered thereof. Let us consider the increasingly common situation in which a new data source is identified to improve or refurbish an official statistical product. What lies ahead preventing a statistical office to access the data? We have empirically identified four sets of issues: legal issues, data characteristics, access conditions, and business decisions.

\subsection{Legal issues}
\label{sub:LegalIssues}
Legal issues constitute apparently the most evident obstacle for a statistical office to access new digital data. It is relevant to underline that access to administrative data has been explicitly included in the main European regulation behind European statistics \citep{EC09a}. In this sense, some countries have already introduced changes in their national regulations to explicitly include these new data sources in their Statistical Acts \citep[see e.g.][]{LoiiNum16a}.\\

Certainly, very deep legal discussions can be initiated around the interpretation and scope of the different entangled regulations in the international and national legal systems but, in our opinion, all boil down to three factors: (i) Statistical Acts, (ii) specific data source regulations, and (iii) general personal data and privacy protection regulations. Regarding Statistical Acts, two main considerations are to be taken into account. On the one hand, by and large these regulations provide legal support for statistical offices to request data to different social agents. On the other hand, more rarely, these regulations also establish legal obligations for these agents to provide the requested data resorting to sanctions in case of nonresponse \citep{LFEP89a}. Regarding data sources such as mobile network data, financial transaction data, online databases, etc.\ there exist commonly specific regulations protecting these data restricting their use only for their specific purposes (telecommunication, finance, online transactions, etc.). These regulations may pose unsolved conflicts with the preceding Statistical Acts. Besides, personal data and privacy protection regulations, whose implementation is usually enacted through Data Protection Agencies, increases the degree of complexity since exceptions for statistical purposes do not explicitly clarify the type of data source to be used for the production of official statistics.\\

When requesting sustainable access in time, all these issues must be surmounted having in mind the perspectives of statistical offices, data holders, and statistical units (citizens and business units). Simultaneously (i) legal support for statistical offices must be clearly stated, (ii) data holders must be also legally supported in providing data, especially about third people (statistical units), and (iii) privacy and confidentiality of all social agents' data must be guaranteed by Law and in practice. Needless to say, Law must be an instrument to preserve rights and establish legal support for all members of society.

\subsection{Data characteristics}
\label{sub:DataChar}

Data ecosystems for new data sources are highly complex and of very different nature. For example, telco data are generated in a complex cellular telecommunication network for many different internal technical and business purposes. Accessing data for statistical purposes implicitly implies the identification of those subsets of data needed for statistical production. Not every piece of data is useful for statistical purposes. Moreover, raw data are not useful for these purposes and they need some preprocessing. Even worse, raw digital data have an unattainable volume for usual production standards at statistical offices and require technical assistance by telco engineers. Thus, some form of preprocessed or even intermediate data may be instead required, but then details about this data processing or intermediate aggregating step need to be shared for later official statistical processing.\\

All in all, the characteristics of new data for the production of official statistics strongly compel the collaboration with data holders. This is completely novel for statistical offices.

\subsection{Access conditions}
\label{sub:AccessCond}

As a result of the complexity behind new data sources, one of the considered options to use this data for statistical purposes is the in-situ access, thus avoiding the risk that data leaves the information systems of data holders. This possibility alleviates the privacy and confidentiality issues, but the operational aspect must then be tackled, since the statistical office will have to access somehow these private information systems. A second option may be to transmit the data from the data holders' premises to the statistical offices' information systems. No access to the private information systems is needed but privacy and confidentiality issues must then be solved in advance, both from the legal and the operational points of view. Finally, a trusted third party may enter the scene who will receive the data from the data holders and then, possibly after some preprocessing, will transmit them to the statistical office. The confidentiality and privacy issue remains open and part of the official statistical production process is further delegated.\\

A second condition comes from the exclusivity for statistical offices to access and use these data. Should there been more social agents requesting access and use of these data sources (e.g.\ other public agencies, ministries, international organizations, etc.), the access conditions from the data holders' point of view would be extremely complex. This raises a natural enquiry about the potential social leading role of statistical offices in making this data available for public good.\\

A third condition revolves around the issue of intellectual property rights and/or industrial secrecy requirements. Accessing these data sources usually entails core industrial process for the data holders, who rightfully wants to protect their know-how from their competitors. Statistical offices must not disrupt the market competence by leaking this information from one agent to another. Guarantees must be offered and fixed in this sense.\\

Fourthly, new data sources will be more efficient when combined among them and with administrative and survey data. Furthermore, in a collaborating environment with data holders it seems naturally to consider the choice to share this data integration (e.g.\ considering this intermediate output as a new statistical product). Operational aspects of this data integration step (especially regarding statistical disclosure control) must be tackled (e.g.\ with secure multiparty computation techniques \citep{ZhaZhaZhaCheGaoLiTan19a}; see also section \ref{sub:PushOutCompParad}).\\

Finally, as partially mentioned above, costs associated to data retrieval, access, and/or processing brought by the complexity of these data sources must be also taken into account. Occasionally this issue does not arise when collaborating for research and for one-shot studies, but it stands as an issue for the long term data provision for standard production. Let us remind the principle 1 of the UN Principles for Access to Data for Official Statistics \citep{UNGWG16a}, where this data provision is called upon free of charge and on a voluntary basis. However, principle 6 explicitly states the \textquotedblleft [t]he cost and effort of providing data access, including possible pre-processing, must be reasonable compared to the expected public benefit of the official statistics envisaged\textquotedblright. Moreover, this is complemented by principle 3 stating that \textquotedblleft [w]hen data is collected from private organizations for the purpose of producing official statistics, the fairness of the distribution of the burden across the organizations has to be considered, in order to guarantee a level playing field\textquotedblright. Thus, these principles arise as pertinent. However, the issue of the cost is extremely intricate. Firstly, the essential principle of Official Statistics by which data provision for these purposes must be made completely free of charge must be respected. Yet, the costs associated to data extraction and data handling for statistical purposes need a careful assessment and this depends very sensitively on the concrete situation of the data holders. Different details need consideration: staff time in data processing, hardware computing time, hardware buy and deployment (if necessary), software development or licenses (if necessary), \dots In addition, the compensation for these costs may be given shape in different ways, from a direct payment to an implicit contribution to a long-term collaboration partnership. In any case, notice that this compensation of costs should not be for the data themselves, but for the data extraction and data handling. Data must be granted access free of charge. Furthermore, if several data holders are at stake for the same data source, equal treatment must be procured for each of them. This is a wholly new social scenario for the production of official statistics.

\subsection{Business decisions}
\label{sub:BusDec}
Apart from the preceding factors, also apparently potential conflicts of interest and risk assessments can advice decision-makers in private organizations not to establish partnerships with statistical offices. The conflicts of interest may arise because of the perception of a potential collision in the target markets between statistical offices and private data holders/statistical producers. Our view is that this is only apparent, that statistical products for the public good considered in National Statistical Plans are of limited profit for private producers, and that in potentially intersecting insights a collaboration will increase the value of all products. Furthermore, corporate social responsibility and activities for social good naturally invite private organizations to set up this public-private collaboration broadening the scope of their activities to increase  the economic and social value of their data, to contribute to the development of national data strategies, and to support policy making more accordingly to their information needs.\\

All in all, access and use of new data sources depend on a highly entangled set of challenging factors for many public and private organizations but offering an extraordinary opportunity to contribute to the production and dissemination of information in the present digital society. Statistical offices should strive to reshape their role to become an active actor in this new scenario.

\section{Methodology}
\label{sec:Method}

As stated in section \ref{sec:Data}, the lack of statistical metadata of new data sources and having data generated before planning and design necessarily impinge directly on the core of traditional survey methodology, especially (but not only) on the limitation in the use of sampling designs for these new data sources. This means that an official statistician accessing a new data source cannot resort to the tools in the traditional (current, indeed) production framework to produce a new statistical output. This does not mean whatsoever that there do not exist statistical techniques to process and analyse this new data. Indeed, there exists a great deal of statistical methods \citep[see e.g.][]{Chapman20a}. We just lack a new extended production framework to cover methodological needs in every statistical domain for each new data source.\\

We shall focus in this section on key methodological aspects in the production of official statistics and share some reflections on the new methods.

\subsection{Key methodological aspects: representativeness, bias, and inference}
\label{sub:RepresBiasInference}

There exist key concepts in traditional survey methodology such as sample representativeness, bias, and inference, which should be assessed in the light of the new types of data. Certainly, survey methodology is limited with new data sources, but it offers a template mirror for a new refurbished production framework to look at: it provides modular statistical solutions for a diversity of different methodological needs along the statistical process in all statistical domains (sample selection, record linkage, editing, imputation, weight calibration, variance estimation, statistical disclosure control, \dots). Furthermore, the connection between collected samples and target populations is firmly rooted on scientific grounds using design-based inference.\\

When considering an inference method other than sampling strategies (sampling designs together with asymptotically unbiased linear estimators), many official statisticians immediately react alluding to sample representativeness. This combination of sampling designs and linear estimators is indeed in the DNA of official statisticians and some first explorations of statistical methods to face this inferential challenge still resemble these sampling strategies \citep{BerLehReiDiCKar18a}. In our view, the introduction of new methods should come with an address on these key concepts (sample representativeness, bias, etc.).\\

To grasp the differences in these concepts in the statistical methods for survey data and for new data sources, we shall shortly include our view on the origin of the strength felt by official statisticians around these concepts in the traditional production framework. As T.M.F.\ \citet{Smi76a} already pointed out, the design-based inference seminally introduced by J.\ \citet{Ney34a} allows the statistician to make inferences about the population \textbf{regardless of its structure}. Also in our view, this is the essential trait of design-based methodology in Official Statistics over other alternatives, in particular, over model-based inference. As M.\ \citet{Han87a} already remarked, statistical models may provide more accurate estimates \textbf{if the model is correct}, thus clearly showing the dependence of the final results on our a priori hypotheses about the population in model-based settings. Sampling designs free the official statistician to make hypotheses sometimes difficult to justify and to communicate.\\

This essential trait appears in the statistical methodology under the use of (asymptotically) design-unbiased linear estimators of the form $\widehat{T}=\sum_{k\in s}\omega_{ks}y_{k}$, where $s$ denotes the sample, $\omega_{ks}$ are the so-called sampling weights (possibly dependent on the sample $s$) and $y$ stands for the target variable to estimate the population total $Y=\sum_{k\in U}y_{k}$. A number of techniques does exist to deal with diverse circumstances regarding both the imperfect data collection and data processing procedures so that non-sampling errors are duly dealt with \citep{LesKal92a, SarLun05a}. These techniques lead us to the appropriate sampling weights $\omega_{ks}(\mathbf{x})$ usually dependent on auxiliary variables $\mathbf{x}$. Sampling weights are also present in the construction of the variance estimates and thus of confidence intervals for the estimates.\\

The interpretation of a sampling weight $\omega_{ks}(\mathbf{x})$ is extensively accepted as providing the number of statistical units in the population $U$ represented by unit $k$ in the sample $s$, thus settling the notion of representativeness on apparently firm grounds. This combination of sampling designs and linear estimators, complemented with this interpretation of sampling weights, stands up as a robust defensive argument against any attempt to use new statistical methodology with digital sources. Indeed, one of the first rightful questions when facing the use of digital data is how data represent the target population. With many new digital sources (mobile network data, web-scraped data, financial transaction data, \dots) the question is clearly meaningful.\\

However, before trying to give due response with new methodology, we believe that it is of utmost relevance to be aware of the limitations of the sampling design methodology in the inference exercise linking sampled data and target populations. This will help producers and stakeholders be conscious about changes brought by new methodological proposals and view the  challenges in the appropriate perspective.\\

Firstly, the notion of representativeness is slippery business. This concept was already analyzed by \citet{KruMos79a, KruMos79b, KruMos79c, KruMos80a} in this line. Surprisingly enough, a mathematical definition in classical and modern textbooks is not found, providing \citet{Bet09a} an exception in terms of a distance between the empirical distributions of a target variable in the sample and in the target population. Obviously, this definition comes with very difficult practical implementation (we would need to know the population distribution). Nonetheless, this has not been an obstacle for the extended use of the concept of representativeness even in a dangerous way. From time to time, one can hear that the construction of linear estimators is undertaken upon the basis of being $\omega_{ks}(\mathbf{x})$ the number of population units represented by the sampled unit $k$, thus amounting $\omega_{ks}(\mathbf{x})\cdot y_{k}$ to the part of the population aggregate accounted for by unit $k$ in the sample $s$, finally being $\sum_{k\in s}\omega_{ks}\cdot y_{k}$ the total population aggregate to estimate. A strong resistance is partially perceived in Official Statistics against any other technique not providing some similar clear-cut reasoning accounting for the representativeness of the sample. This argument is indeed behind the restriction upon sampling weights for them not to be lesser than $1$ (interpreted as a unit not representing itself) or for them to be positive in sampling weight calibration procedures (see e.g.\ \citet{Sar07a}). In our view, the interpretation of a unit $k$ in a sample as representing $\omega_{ks}$ units in the population can be impossible to justify even in such a simple example as a Bernoulli sampling design of probability $\pi=\frac{1}{2}$ in a finite population of size $N = 3$: if, e.g., $s=\{1, 2\}$, how should we understand that these two units represent $4$ population units?\\

Ultimately, the goal of an estimation procedure is to provide an estimate as close as possible to the real unknown target quantity together with a measure of the accuracy. The concept of mean square error, and its decomposition in bias and variance components \citep{Gro89a}, is essential here. Estimators with a lower mean square error guarantee a high-quality estimation. No mention to representativeness is needed. Furthermore, not even the requirement of exact unbiasedness is rigorously justified:  compare the estimation of a population mean using an expansion (Horvitz-Thompson) estimator and using the H\'{a}jek estimator \citep{Haj81a}.\\

The randomization approach does allow the statistician not to make prior hypotheses on the structure of the population to conduct inferences, i.e.\ the confidence intervals and point estimates are valid for any structure of the population. But this does not necessarily entail that the estimator must be necessarily linear. Given a sample $s$ randomly selected according to a sampling design $p(\cdot)$ and the values $\mathbf{y}$ of the target variable, a general estimator is any function $T=T(s, \mathbf{y})$, being linear estimators a specific family thereof \citep{HedSin91a}. Thus, what prevents us to use more complex functions provided we search for low mean square error? Apparently nothing. A linear estimator may be viewed as a homogeneous first-order approximation to an estimator $T(s,\mathbf{y})$ such as $T(s,\mathbf{y})\approx \sum_{k\in}\omega_{ks}y_{k}$, but why not a second-order approximation $T(s,\mathbf{y})\approx \sum_{k\in}\omega_{ks}y_{k} + \sum_{k,l\in s}\omega_{kls}y_{k}y_{l}?$ Or even a complete series expansion $T(s,\mathbf{y})\approx \sum_{p=1}^{\infty}\sum_{k_{1},\dots, k_{p}\in s}\omega_{k_{1}\dots k_{p} s}\cdot y_{k_{1}}\dots y_{k_{p}}$ (see e.g. \citet{LehVei98a})?\\

However, the multivariate character of the estimation exercise at statistical offices provides a new ingredient shoring up the idea of representativeness, especially through the concept of sampling weight. Given the public dimension of Official Statistics usually disseminated in numerous tables, \textbf{numerical consistency} (not just statistical consistency) is strongly requested on all disseminated tables, even among different statistical programs. For example, if a table with smoking habits is disseminated broken down by gender and another table with eating habits is also disseminated broken down by gender, the number of total women and men inferred from both tables must be \textbf{exactly} equal. Not only is this restriction of numerical consistency demanded among all disseminated statistics in a survey but also among statistics of different surveys, especially for core variables such as gender, age, or nationality. Linear estimators can be made easily fulfilled this restriction by forcing the so-called \emph{multipurpose property of sampling weights} \citep{Sar07a}. This entails that the same sampling weight $\omega_{ks}$ is used for any population quantity to estimate in a given survey. For inter-survey consistency, sometimes the calibration of sampling weights is (dangerously) used. This elementarily guarantees the numerical consistency of all marginal quantities in disseminated tables.\\

Notice, however, that this property has to be forced. Indeed, the different techniques to deal with non-sampling errors (e.g. non-response or measurement errors) rely on auxiliary information $\mathbf{x}$ so that sampling weights $\omega_{ks}$ are functions of these auxiliary covariates $\omega_{ks}=\omega_{ks}(\mathbf{x})$. Forcing the multipurpose property amounts to forcing the same behaviour in terms of non-response, measurement errors, etc.\ (thus in terms of social desirability or satisficing response mechanisms) regarding \emph{all} target variables in the survey. Apparently it would be more rigorous to adjust the estimators for non-sampling errors on a separate basis looking only for a \emph{statistical consistency} among marginal quantities. However, this is much harder to explain in the dissemination phase and traditionally the former option is prioritized paving the way for the representativeness discourse (now every sampled unit is thought to \textquotedblleft truly\textquotedblright\ represent $\omega_{ks}$ population units).\\

Secondly, sampling designs are thought of as a life jacket against model misspecification. For example, even not having a truly linear model between the target variable $y$ and covariates $\mathbf{x}$, the GREG estimator is still asymptotically unbiased \citep{SarSweWre92a}. But (asymptotical) design-unbiasedness does not guarantee a high-quality estimate. A well-known example can be found in Basu's elephants story \citep{Bas71a}. Apart from implications in the inferential paradigm, this story clearly shows how a poor sampling design drives us to a poor estimate, even using exactly design-unbiased estimators. A design-based estimate is good \textbf{if the sampling design is correct}.\\

Finally, as already well-known in small area estimation techniques \citep{RaoMol15a} and as R.\ \citet{Lit12a} called \emph{inferential schizophrenia}, sampling designs cannot provide a full-fledged inferential solution for all possible sample sizes out of a finite population. Traditional estimates based on sampling designs show their limitations when the size of the sample for population domains begins to decrease dramatically. With new digital data one expects to avoid this problem by having plenty of data, but in the same line one of the expected benefits of the new data sources is to provide information at an unprecedented space and time scale. So, the problem may still remain in rare population cells.\\

In our view, thus, we must keep the spirit for representativeness in an abstract or diffuse way, for lack of bias, and for low variances, as in traditional survey methodology. But we should avoid some restrictive misconceptions and open the door to find solutions in the quest for accurate statistics with new data sources. There exist multiple statistical methods which should be identified to conform a more general statistical production framework. Probability theory can still provide a firm connection between collected data sets and target populations of interest.

\subsection{New production framework: probability theory, learning, and artificial intelligence}
\label{sub:NewProdFrwk}
We do not dare to provide an enumeration of statistical methods conforming the new production framework. Much further empirical exploration and analysis of the new data sources are needed to furnish a solid production framework and this will take time. However, some ideas can already be envisaged. The impossibility of using sampling designs necessarily makes us resort to statistical models, which essentially amounts to the conception of data as realizations of \textbf{random variables} \citep{LehCas98a}. As stated above, notice that this was not the case for the inferential step in survey methodology (although it was supplementarily made for other production steps as e.g.\ imputation).\\

The consideration of random variables as a central element brings immediately into scene the distinction between the enumerative and analytical aims of official statistical production \citep{Dem50a}. Let us use an adapted version of exercise 1 in page 254 of the book by \citet{Dem50a}. Consider an industrial machine producing bolts according to a given set of technical specifications (geometrical form, temperature resistance, weight, etc.). These bolts are packed into boxes of a fixed capacity (say, $N$ bolts) which are then distributed for retail trade. We distinguish two statistically different (though related) questions about this situation. On the one hand, we may be interested to know the number of defective bolts in each box. On the other hand, we may be interested to know the rate of production of defective bolts by the machine. Both questions are meaningful. The retailer will naturally be interested in the former question whereas the machine owner will also be interested in the latter. Statistically, the former question amounts to the problem of estimation in a finite population \citep{CasSarWre77a} while the latter is a classical inference problem \citep{CasBer02a}. Indeed, the concept of sample in both situations is different (see the definition of sample by \citet{CasSarWre77a} for a finite population setting and that by \citet{CasBer02a} for an inference problem). Notice that the use of inferential samples is not extraneous to the estimation problem in finite populations. The prediction-based approach to finite-population estimation \citep{ValDorRoy00a, ChaCla12a} already makes use of target variables as random variables.  In traditional official statistical production, the former sort of question is solved (number of unemployed people, of domestic tourists, of hectares of wheat crop, etc.). With new data sources and the need to consider data values as realizations of random variables, should Official Statistics begin considering also the new questions?\\

In this line, there already exists an important venue of Statistics and Computer Science research which Official Statistics, in our view, should incorporate in the statistical outputs included in National Statistical Plans. Traditionally, the focus of the estimation problem in finite population has been totals of variables providing aggregate information for a given population of units broken down into different dissemination population cells. The wealth of new digital data opens up the possibility to investigate the \textbf{interaction} between those population units, i.e.\ to investigate \textbf{networks}. Indeed, a recent discipline has emerged focusing on this feature of reality (see \citet{Bar08a} and multiple references therein). Aspects of society with public interest regarding the interaction of population units should be in the focus of production activities in statistical offices. New questions as the representativeness of interactions in a given data set with respect to a target population arise as a new methodological challenge in Official Statistics.\\

A closer look at the mathematical elements behind this network science will reveal the versatile use of graph theory \citep{Bol02a, Ste10a} to cope with complexity. As a matter of fact, the combination of probability theory and graph theory is a powerful choice to process and analyse large amounts of data. Probabilistic graphical models \citep{KolFri09a}, in our view, should be part of the methodological tools to produce official statistics with new data sources. They provide an adaptable framework to deal with many situations such as speech and pattern recognition, information extraction, medical diagnosis, genetics and genomics, computer vision and robotics in general, \dots This is already bringing a new set of statistical and learning techniques into production.\\

This immediately takes us to machine learning and artificial intelligence techniques. In this regard, we should distinguish between the inferential step connecting data and target populations and the rest of production steps. Many tasks, old and new, can be envisaged as incorporating these recent techniques to gain efficiency. Traditional activities such as data collection, coding, editing, imputation, etc.\ can be presumably improved with random forests, support vector machines, neural networks, natural language processing, etc. New activities such as pattern and image recognition, record deduplication, [\dots] will also be conducted with these new techniques. Further research and innovation must be carried out in this line.\\

For the inferential step, however, we cannot see these new techniques as a definitive improvement. Our reasoning goes as follows. An essential ingredient in machine learning and artificial intelligence is \textbf{experience} \citep{GooBenCou16a}, i.e.\ the accumulation of past data from which the machine or the intelligent agent will learn. Learning to make inferences for a target population entails that we know and accumulate the ground truth so that algorithms can be trained and tested. The ground truth for a target population is never known. Thus, the inference step must receive the same attention as in traditional production. There may be situations in which the wealth and nature of digital data may bring the case where the whole target population is sampled (e.g.\ a whole national territory can be covered by satellite images to measure the extensions of crops), but even in those cases the treatment of non-sampling errors must be taken into account (as already envisaged by \citet{Yat65a}).\\

This incorporation of new techniques from fields like machine learning and artificial intelligence entails a necessity to set up a common vocabulary and understanding of many related concepts in these disciplines and in traditional statistical production. Let us focus, e.g., on the notion of bias. This arises once and another both in machine learning and in estimation theory. In traditional finite population estimation, the bias $\mathbb{B}(\widehat{Y})$ of an estimator $\widehat{Y}$ of a population total $Y$ is defined with respect to the sampling design $p(\cdot)$ as $\mathbb{B}(\widehat{Y})=\mathbb{E}_{p}(\widehat{Y})-Y$, which amounts basically to an expectation value \emph{over all possible samples}. In survey methodology, estimators are (asymptotically) unbiased by construction. This notion of bias is not to be confused with the difference between the true population total $Y$ and an estimate from the selected sample $\widehat{Y}(s)$. This estimate error $\widehat{Y}(s) - Y$ is never known and can be non-zero even for exactly unbiased estimators. When the prediction approach is assumed and the population total is also considered a random variable, the concept of (prediction) bias is slightly different: $\mathbb{B}(\widehat{Y})=\mathbb{E}_{m}(\widehat{Y}-Y)$, where $m$ stands for the data model. These notions of population bias are not to be confused with the measurement error $y_{k}^{\textrm{obs}}-y_{k}^{(0)}$, where $y_{k}^{\textrm{obs}}$ stands for the raw value observed in the questionnaire and $y_{k}^{(0)}$ stands for the true value for unit $k$ of variable $y$. Indeed, in statistical learning this is very often referred to as bias. We model variable $y$, indeed. An effort to build a precise terminology when new techniques are used is needed in order to assure a common understanding of the mathematical concepts at stake. Another example comes from the reference to linear regression as a \textquotedblleft machine learning algorithm\textquotedblright\ \citep{GooBenCou16a}. New techniques bring new useful perspectives even in the traditional process but the community of official statistics producers must be sure that communication barriers do not arise.\\

Finally, apart from machine learning and artificial intelligence and in connection with different aspects of data access and data use already mentioned in section \ref{sec:Access}, we must make a special mention to data collection and data integration. New digital data per se will provide individually a high value to official statistical products but arguably it is the integration and combination of them together with survey and administrative sources which will boost the scope of future statistical products. At this moment, this integration and combination is thought to be potentially conducted only with no disclosure of each integrated database. This drives us necessarily to cryptology and the incorporation of cryptosystems in the production of official statistics. Notice, however, that this does not substitute the statistical disclosure control upon final outputs, which must still be conducted. Now it is also at the input of the statistical process where data values are not to be undisclosed. The cryptosystem must be able to carry out complex statistical processing in an undisclosed way. A lot of research in this line is needed. \\

All in all, new methods are to be incorporated with the new data sources, many of them already existing in other disciplines. The challenge is to furnish a new production framework. New data and new methods bring necessarily considerations for quality, for the technological environment, and for staff capabilities and management within statistical offices.

\section{Quality}
\label{sec:Qual}
Quality has been a distinguishing feature of official statistical production for many decades and lots of efforts have been traditionally devoted to reach high-quality standards in survey-data-based publicly disseminated statistical products. With new data sources these high-quality standards must be also pursued.\\

We identify key notions in current quality systems in Official Statistics and try to understand how they are to be affected by the nature of the new data sources and the new needs in statistical methodology. We underline three important notions. Firstly, the concept of quality in Official Statistics evolved from the exclusive focus on accuracy to the present multidimensional conception in terms of (i) relevance, (ii) accuracy and reliability, (iii) timeliness and punctuality, (iv) coherence and comparability,  and (v) accessibility and clarity \citep{ESS14a}. Current quality assurance frameworks in national and international statistical systems implement this multidimensional concept of quality (or slight variants thereof). Will new quality dimensions be needed? Will existing quality dimensions be unnecessary? Secondly, a statistical product is understood to have a high-quality standard if it has been produced by a high-quality statistical process. How will the changes in the statistical process affect quality? Thirdly, quality is mainly conceived of as \textquotedblleft fit for purposes\textquotedblright \citep{Eur20b}. How will statistical products based on new data sources be fit for purposes? Certainly, these are not orthogonal unrelated notions, but they can jointly offer a wide overview of the main quality issues.\\

\subsection{Quality dimensions, briefly revisited}
\label{sec:QualDim}

Regarding the quality dimensions, we do not foresee a need to reconsider the current five-dimensional conception mentioned above. Already with traditional data, alternative more complex multidimensional views of data quality could already be found in the literature \citep[see e.g.][and multiple references therein]{WanWan96a}. In our view, the nature of new data sources will certainly require a revision of the existing dimensions, especially the conceptualization and computation of some quality indicators, but not the suppression and/or consideration of new dimensions. Let us consider as an immediate relevant example the consequences of using model-based inference (possibly deeply integrated in complex machine learning or artificial intelligence algorithms). Parameter setting, model choice, and any form of prior hypothesis regarding the model construction must be clearly assessed and communicated. This ingredient impinging on accuracy, comparability, accessibility, and clarity gains in relevance with new data sources.\\

We comment very briefly on the aforementioned quality dimensions:

\begin{itemize}
\item Relevance is essentially an address to current and potential statistical needs of users. This dimension is deeply entangled with our third question regarding being fit for purpose. We will deal with this dimension more extensively below.
\item Accuracy is directly impinged by the new methodological scenario. Inference cannot be design-based with new data sources, thus model-based estimates will gain more presence. Furthermore, since these new data sources come mostly from event-register systems, the usual reasoning on target units and target variables is not directly applicable, thus reducing the validity of the usual classification of errors (sampling, coverage, non-response, measurement, processing). These errors are severely survey-oriented and despite the possibility of more generic readings of current definitions we find it necessary to undergo a detailed revision. Let us consider a hypothetical situation in which a statistical office has access to all call detail records (CDRs) in a country for a given time period of analysis to estimate present population counts. These network events are generated by an active usage of a mobile device. Discard children, very elderly people, imprisoned people, severely deprived homeless people, and any rather evident non-subscriber of these mobile telecommunication services. Can all CDRs be considered a sample with respect to our (remaining) target population? There is no more CDR data, however we cannot be sure that all target individuals can be considered included in the dataset. Indeed, there is no enumeration of the target population and the \textquotedblleft error[s] [\dots] which cannot be attributed to sampling fluctuations\textquotedblright \citep{ESS14a} cannot be clearly identified. The line distinguishing coverage and sampling errors becomes thinner (as a matter of fact, the concept of frame population in this new setting loses its meaning).\\

    Reliability and the corresponding plan of revisions can still be considered under the same approach as in traditional data sources, only potentially affected by both the higher degree of breakdown and availability of data. When dissemination cells are very small and publicly released more frequently, the variability of estimates are expected to be much higher. Thus, an assessment to discern between random fluctuations because of small-sized samples and fluctuations because of real effects (e.g.\ population counts attending music festivals or sport events) is needed. The plan of revisions should be accommodated to the chosen degree of breakdown in the dissemination stage.

\item Timeliness arises as one of the most clearly improved quality dimensions when incorporating new data sources. Indeed, with digital sources even (quasi) real-time estimates may be an important novelty. However, this is intimately connected to the design and implementation of the new statistical production process and the relationship with data holders. Real-time estimates entail real-time access and processing, which is usually highly disruptive and requires a higher investment on the data retrieval and data preprocessing stages, presumably on data holders' premises. Therefore, guarantees (both legal and technical) for access sustained in the long term must be provided. Once timeliness can be improved, new output release calendars can be considered in legal regulations for each statistical domain, thus binding statistical offices to disseminate final products with the same punctuality standards.\\
\item The role of coherence and comparability is to be reinforced with new data sources. The reconciliation among other sources, other statistical domains, and other time-frequency statistics is now more critical. It is not only that the data deluge will allow statistical offices to reuse the same source to produce different statistics for different statistical domains (e.g.\ financial transactional data for retail trade statistics, for tourism statistics, \dots), but also that different sources will possibly lead to estimates for the same phenomenon (e.g.\ unmanned aerial images, satellite images, administrative data, survey data for agriculture). This is naturally connected also to comparability, since statistical products must still be comparable between geographical areas and over time. The criticality is intensified because the wealth of statistical methods and algorithms potentially applicable on the same data can lead to multiple different results where the comparison is not immediate. This demands a closer collaboration in statistical methodology in the international community.
\item Accessibility and clarity in relation to users is essential (e.g.\ to the point expressed above of strongly nailing the non-mathematical notion of representativeness in the world of Official Statistics). The challenge raised by the wealth of statistical methods and machine-learning algorithms to solve a given estimation problem stands now as an extraordinary exercise in communication strategy and policy. Furthermore, this communication strategy and policy should not only embrace but also get deeply entangled with the access and use of the new data sources. The promotion of statistical literacy will need to be strengthened.
\end{itemize}

\subsection{Process quality}
\label{sub:ProcQual}
Changes in the process will certainly be needed according to the new methodological ingredients mentioned in section \ref{sec:Method}. As a matter of fact, the implementation of new data sources in the production of some official statistics is already bringing the need for new business functions such as trust management, communication management, visual analyses, \dots \citep{WPF1.19a, KuoLoi20a}. However, in our view the farthest-reaching element will come from the need to include data holders as active actors in the early (and not so early) stages of the production process. This will especially affect those deeply technology-dependent data sources with a clear data preprocessing need for statistical purposes. In other words, data holders have changed their roles from just input data providers, either through electronic or paper questionnaires, to also data wrangler for further statistical processing.\\

Being official statistics a public good, it seems natural to request that this participation of data holders will need to be reflected in quality assurance frameworks to assess their impact on final products. In our view, this entails far-reaching consequences and strongly imposes conditions on the partnerships between statistical offices and data holders. These conditions are two-fold, since restrictions both for the public and private sector must be observed. For example, the statistical methodology driving us from the raw data to the final product must be openly disseminated, communicated and available to all stakeholders as an integral elemental of the statistical production metadata system. Furthermore, to guarantee coherence and comparability it seems logical to share this statistical methodology among different data holders, i.e.\ in any preprocessing stage. However, guarantees must also be provided to avoid sensitive information leakage among different agents in the private sector, especially in highly competitive markets. statistical offices cannot become malicious vectors of industrial secrecies and know-hows endangering an increasing economic sector based on data generation and data analytics. Another example comes from data sources providing geolocated information (e.g.\ to estimate population counts of diverse nature). Current data technologies allow us to reach unprecedented degrees of breakdown (e.g.\ providing data every second minute at postal code geographical level). Freely disseminating population counts at this level of breakdown in a statistical office website would certainly ruin any business initiative to commercialise and/or to foster private agreements to produce statistical products. Partnerships must include formulas of collaboration where both private and public interests not only can, in our opinion, coexist, but even also positively feedback each other.\\

In this line of thought, synthetic data can play a strategic role, even beyond traditional quality dimensions and traditional metadata reporting. In our view, an important aspect of the public-private partnerships with data holders is a deep knowledge of metadata of the new data sources. This would enable statistical offices to generate synthetic data with similar properties to real data. This synthetic data can play a two-fold role. On the one hand, for all data sources, providing synthetic data together with process metadata will enable users and stakeholders to get acquainted with the underlying statistical methodology thus increasing the overall quality in the process. For example, a frame population of synthetic business units can be synthetically created so that the whole process from the sample selection to the final dissemination phase and monitoring can be reproduced. On the other hand, for new data sources with those challenges in access and use reported above, methodological and quality developments as well as software tools can be investigated without incurring on those obstacles with real data. Notice that the utility of this synthetic data will sensitively depend on their similarity with real data, thus demanding a good knowledge of their metadata, i.e.\ calling for a close collaboration with data holders.

\subsection{Relevance}
\label{sub:Relevance}
Relevance is a quality attribute measuring the degree to which statistical information meets the needs of users and stakeholders. Thus, it is intimately related to outputs being \emph{fit for purpose}. Moreover, relevance is one of the key issues in the Bucharest Memorandum \citep{DGINS18a} clearly pointing out the risk for public statistical systems in case of not incorporating new data sources into the production process (among other things).\\

In more mathematical terms, let us view relevance in terms of the nature of statistical outputs and aggregates. Up to current dates, most (if not all) statistical outputs are estimates of population totals $\sum_{k\in U}y_{k}$ or functions of population totals $f(\sum_{k\in U}y_{k}, \sum_{k\in U}z_{k}, \dots)$. They may be the total number of unemployed resident citizens, the number of domestic tourists, the number of employees in an economic sector, etc., but also volume and price indices, rates, and so on. This sort of outputs is basically built using estimates of quantities such as $\sum_{k\in U_{d}}y_{k}$, where $U_{d}$ denotes a population domain and $y_{k}$ stands for the the fixed values of a target variable. In our view, the wealth of data provides now the opportunity to investigate a wider class of indicators. Network science \citep{Bar08a} provides a generic framework to investigate new kinds of target information, in particular, that derived from the interaction between population units. Graph theory stands up as a versatile tool to pursue these ideas. If nodes represent the target population units, edges express the relationship among these population units. An illustrative example can be found in mobile network data, where edges between mobile devices can represent the communication between people and/or with telecommunication services. If the geolocation of this data is also taken into account and they are combined with other data sources (e.g.\ financial transaction data -- also potentially geolocated),  many new possibilities arise to investigate e.g.\ segregation, inequalities in income, access to information and other services, etc. New statistical needs naturally arise. Should statistical offices act reactively waiting for users to express these new needs or should they act proactively searching for new forms of information, new indicators, and new aggregates? In our view, innovation activities and collaboration with research centres and universities should be strengthened to promote proactive initiatives.

\section{Information technologies}
\label{sec:IT}

\subsection{IT for survey and administrative data}
\label{sub:ITsurveyAdmin}

The very fast evolution of the information technologies has changed our lives. Nowadays, almost every human activity leaves a digital footprint: from searching information on Internet using a search engine to using a mobile phone for a simple call or paying a product with a credit card, the traces of these activities are stored somewhere in a digital database. Accordingly, these enormous quantities of data draw the attention of statisticians who started to consider their potential for computing new indicators. The distinct characteristics of these new data sources that were emphasized in the previous sections also changed the IT tools needed to tackle with them. While using the classical survey data to produce statistical outputs doesn't raise special computational problems, collecting and processing new types of data (that are most of the times very big in volume) requires an entire new computing environment as well as new skills for the people that work with them. In this section we will shortly review the computing technologies used in official statistics for dealing with survey data and we will describe the new technologies needed to handle new big data sources. We emphasize that the computing technologies are evolving with an unprecedented speed and what it seems to be now the best solution, in few years could be totally outdated. We will also provide some examples of concrete computing environments used for experimental studies in the official statistics area.\\

The computing technology needed for a specific type of a data source is intrinsically related to the nature of the data source. Survey data are structured data with a reasonable size, properties that make them easy to store with traditional relational databases.  The IT tools used for surveys can be classified according to the specific stage in the production pipeline and for this purpose we will consider the GSBPM as the general framework describing the official statistics production process.\\

Different phases of the statistical production process such as drawing the samples, data editing and imputation, calculation of aggregates, calibration of the sampling weights, seasonal adjustments of the time series, performing statistical matching or record linkage use specialized software routines, most of the time developed in-house by some statistical agencies and then shared with the rest of the statistical community, that are implemented using commercial products like SAS, SPSS, Stata or open source software like R or Python.\\

While in the past most of the official statistics bureaus where strongly dependent on a commercial software packages like SAS or Stata for example, nowadays we are witnessing a major change in this field. The benefits of the open source software were reconsidered by the official statistics organizations and more and more software packages are now ported to the R or Python ecosystems \citep{mark}.\\

The data collection stage in the production pipeline requires specialized software. Even if the paper questionnaires are still in use in several countries around the world, the main trend today is to collect survey data using electronic questionnaires \citep{Bet09b, Sal19} by either CAPI or CAWI method. In both cases, specific software tools are required to design the questionnaires and to effectively collect the data.  We mention here some examples of software tools in this category:
\begin{itemize}
	\item BLAISE \citep{Bla19} is a computer aided interviewing system (CAI) developed by CBS which is currently used worldwide in several fields: from household to business and economic or labor force surveys. According to the official web page of the software (\url{https://www.cbs.nl/en-gb/our-services/blaise-software}) more than 130 countries use this system. It allows statisticians to create multilingual questionnaires that can be deployed on a variety of devices (both desktops and mobile devices), it is supported by all major browsers and operating systems (Windows, Android, iOS) and has a large community of users. More, BLAISE is not only a questionnaire designer and data collection tool but it can also be used in all stages of the data processing.
	\item CSPro (Census and Survey Processing System) is a freely available software framework for designing applications for both data collection and data processing. It is developed by the U.S. Census Bureau and ICF International. The software can be run only on  Windows systems to design data collection applications that can be deployed on devices running Android or Windows OS. It is used by official statistics institutes, international organizations, academic institutions and even by private companies in more than 160 countries (\url{https://census.gov/data/software/cspro.html}).
	\item Survey Solutions \citep{SuSo18a} is a free CAPI, CAWI and CATI software developed by the World Bank for conducting surveys. The software has capabilities for designing the questionnaires, deploying them on mobile devices or on Web servers, collect the data, perform different survey management tasks and it is used in more than 140 countries \citep{SuSo18b}.
\end{itemize}
There are also other software tools for data collection but they are used on a smaller scale being built by statistical offices for their specific needs.\\

\begin{figure}[h]
	\centering
	\includegraphics[width=8cm]{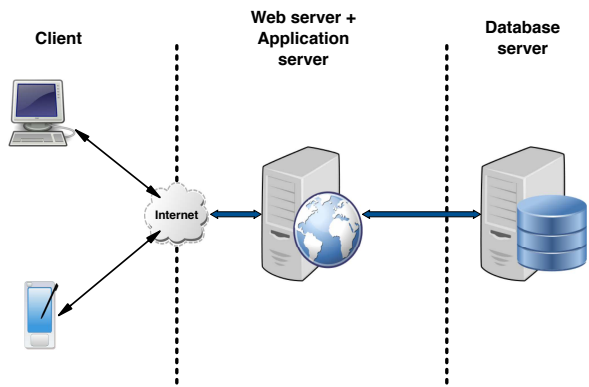}
	\caption{The client-server model of computing}\label{fig:client-server}
\end{figure}

All these tools used in official statistics for data collection are built around a well know technology: the client-server model. Even this model dates from the 1960s and 1970s when the foundations of the ARPANET where built \citep{rfc4, rfc5} it becomes very popular with the appearance and the development of the Web system that transformed the client into the ubiquitous Web browser, making the entire system easier to deploy and maintain. Nowadays there are a plethora of computing technologies supporting this model: Java and .NET platforms, PHP together with a relational database, etc. Figure \ref{fig:client-server} describes the architecture of a typical client-server  application when the client is a Web browser.\\

The client which is usually a browser running on a mobile device or on a desktop loads a questionnaire used to collect the data from households or business units. These data are subject to preliminary validation operations and then they are sent to the server side where a Web server manages the communications via HTTP/HTTPS protocol and an application server implements the logic of the information system. Usually, some advanced data validation procedures are performed before the data are sent to a relational database. From this database, the datasets are retried by the productions units that start the processing stage.\\

The last stages of the production pipeline, i.e. the dissemination of the final aggregates, require also specialized technologies.
Statistical disclosure control (SDC) methods are special techniques with the aim of preserving the confidentiality of the disseminated data to guarantee that no statistical unit can be identified. These methods are implemented in software packages, most of them being in the open source domain. We can mention here \textbf{sdcMicro} \citep{Tpl15} and \textbf{sdcTable} \citep{Mdl19} R packages or \textbf{tauArgus} \citep{Tau14} and \textbf{muArgus} \citep{Mu14} Java programs.\\

Even for disseminating the results on paper, software tools are still needed: starting from the classical office packages which are easy to use by statisticians to more complex tools like \textbf{Latex} that require specific skills, all the paper documents are produced using IT tools. In the digital era the dissemination of the statistical results switched to the Web pages where technologies based on Javascript libraries like \textbf{D3} \citep{d3} or R packages like \textbf{ggplot2} \citep{ggplt16} are widespread.\\

In general, the administrative sources are treated using the same software technologies like survey data with the exception of the data collection step which is not needed in this case.

\subsection{IT for digital data: overview}

The new data sources bring also new information technologies on the stage of official statistics. Accidentally or not, with the beginning of the use of new data sources, a new trend has manifested itself in official statistics: the open source software revolution has also been embraced by the world of official statistics. Two software environments emerged as being suitable for official statistics tasks: R and Python. While Python is considered to be more computationally efficient, R is better suited for statistical purposes: there are R packages for almost every statistical operation, from sampling to data visualisations. In the European Statistical System (ESS) it seems that R has gained ground against Python. Most of the National Statistical Organizations (statistical offices) within the ESS make a transition from old software packages mostly based on commercial solutions to the  R environment \citep{templaus, markrrs}.\\

We mention here only few of the R packages used in statistical offices for different tasks. For drawing surveys samples there are packages like \textbf{sampling} \citep{sampl} that allow not only to use different sampling algorithms but also to calibrate the design weights. \textbf{ReGenesees} package \citep{rgs} developed by ISTAT starts from the \textbf{survey} package \citep{srv} that provides functions to compute totals, means, ratios and quantiles for the survey sample and includes calibration and sampling variance estimation functions. Other R packages used to draw samples with a specific design are \textbf{SamplingStrata} \citep{strat},  \textbf{FS4} \citep{fs4}, \textbf{MAUSS-R} \citep{mauss}.
Visualising and editing the data sets can be performed with \textbf{editrules} \citep{edtr} or \textbf{VIM} \citep{vim}
packages while for selective editing there are packages like \textbf{SeleMix} \citep{slmx}. Imputation can be performed with \textbf{VIM} \citep{vim}, \textbf{mice} \citep{mice} or \textbf{mi} \citep{mi} packages. For time series analyses and seasonal adjustments there are \textbf{x12} \citep{x12} and \textbf{seasonal} \citep{ssn} packages besides the well-known \textbf{JDemetra+} Java software \citep{jdm}. Statistical matching and record linkage is another domain where we can find good quality R packages: \textbf{StatMatch} \citep{stmtch}, \textbf{MatchIt} \citep{mtchit}, \textbf{RecordLinkage} \citep{rlk} and \textbf{RELAIS} \citep{rels}.  Our enumeration is not intended to be exhaustive but to give the reader an image of the capabilities of the R environment for statistical data processing. A comprehensive list of R packages used in official statistics is published at \url{https://github.com/SNStatComp/awesome-official-statistics-software}.

\subsection{New IT tools for data collection}
\label{sub:NewITDataCollect}

The new types of data sources require different technologies for data collection step. If the data sets are to be stored inside NSIs premises, either they are transfered from the data owners using specialized transmission lines, either they are collected using specific technologies. For example, one of the most promising data source is the Internet, or to be more specific, Web sites. There are several technologies that were developed by statistical offices to collect different kind of data (for example prices from online retailers, enterprise characteristics from companies' Web sites, information about job vacancies from specialized portals, etc.) collectively gathered under the term \textit{web scrapping} techniques.\\

\begin{figure}[h]
	\centering
	\includegraphics[width=8cm]{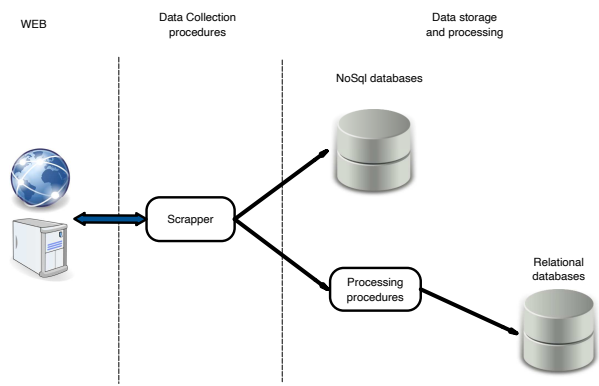}
	\caption{Data collection through web-scraping} \label{fig:web-scrapping}
\end{figure}

In figure  \ref{fig:web-scrapping} we depicted the general organization of such a data collection approach for an IT point of view. There are several solutions used by different statistical offices to implement the main component of this system, called the \textit{Scrapper} in this figure. Some of them are based on R packages, some on Python libraries, others are specific software solutions developed in-house or are based on open source projects  For example, \textbf{rvest} \citep{rvst} is an R package that can be used to scrape data from static \textit{html} pages. It needs an URL and it can gets the entire page or, if the user provides a selector on that page, it gets only the text associated with that selector. The data obtained in this case is text and it is usually stored in a NoSql database or it is processed according to some specific needs and transformed in structured data that is stored in a relational database. Similar packages such as \textbf{scrapeR} \citep{scrp} or \textbf{Rcrawler} \citep{rcrl} can be successfully used for static Web pages.\\

Most of the sites today are actually dynamic and this feature raises some problems when it comes to scrap such pages.  A solution often used to scrape dynamic sites using the R technology is based on the \textbf{RSelenium} \citep{rsln} package which is an R client for Selenium Remote WebDriver. It allows the user to scrape content that is dynamically generated by driving a browser natively, emulating the actions of a real user, and it can be used to automate tasks for several browsers: Firefox, Chrome, Edge, Safari or Internet Explorer. A similar client is also available for Python.\\

Another versatile solution for Web scrapping is the Python \textbf{Scrapy} \citep{scrpy}  which is an application framework that allows users to write Web crawlers that extract structured data from Web sites. An example of a real world applications of this framework in the field of official statistics there are a set of projects developed by ONS \citep{onsscrp, onsscrp2} to collect price data from Internet to compile price indices.\\

Besides these tools, we can also mention in-house software solutions for Web scraping such as the Robot framework developed and used by CBS \citep{cbsrbtf} or a solution based on the Apache Nutch technology \citep{nutch} used by ISTAT for an internal project regarding collection of enterprise characteristics from Web sites.\\

\subsection{New IT tools for data processing}
\label{sub:NewITDataProcess}

The processing step of new data sources takes into account their specificity, esspecially the very large volume. This requires either to use parallel programming paradigms inside an ecosystem like R or Python, or to use dedicated IT architectures.\\

\begin{figure}[h]
	\centering
	\includegraphics[width=8cm]{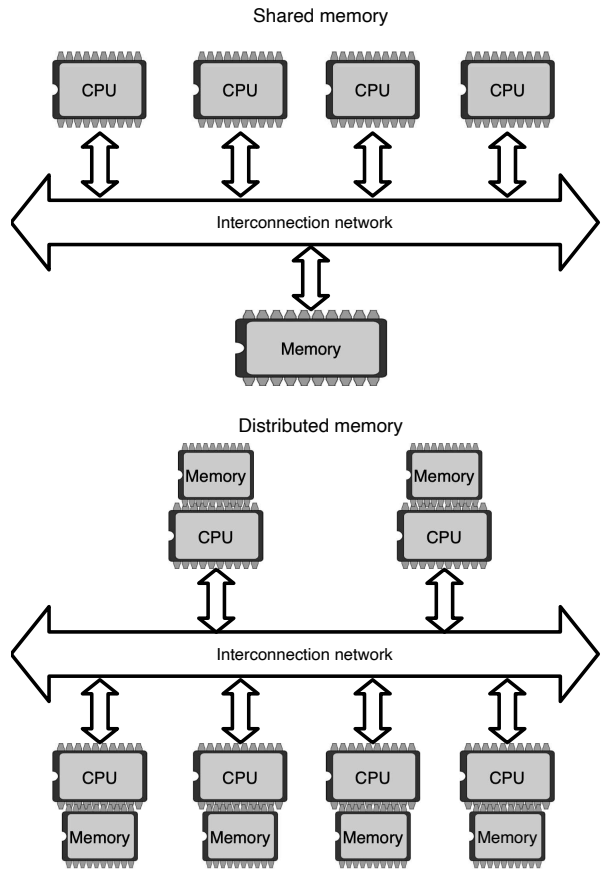}
	\caption{Shared memory versus distributed memory parallelism} \label{fig:parallel}
\end{figure}

The simplest solution for processing large data sets is to use the parallel programming features incorporated in software systems like R or Python. They make use of the multicore any many core architectures of the current computing systems. Two paradigms emerged in this area: shared  and distributed memory architectures. These two models are depicted in figure \ref{fig:parallel}.\\

In the first approach a set of CPUs are interconnected with a single shared memory and all of them have access to this common memory. All modern processors are multicore and they are based on an architecture very similar with the one presented in upper part of the figure \ref{fig:parallel}. However, there is an important limitation of this type of architecture of a computing system: all CPUs are competing for access to the same memory. This severely limits the performance of a computing system even there are solutions that alleviates to some extent this problem.\\

In the second approach several CPUs that have their own memory are interconnected, forming thus a distributed-memory computer. This solution scales up to thousands of CPUs or even more. Tasks can be run in parallel by different CPUs having the necessary data in their own memory, avoiding thus the memory contention problem. At certain steps of the processing algorithms it may be necessary for the CPUs to exchange data between them via the interconnection network or to synchronize themselves.\\

Both aproaches are used for statistical data processing. In the following  we will use R examples, but similar technologies are available for Python too. Parallel computing in the shared memory architecture can be implemented in R  via compiled extensions that rely on specific compiler support: OpenMP \citep{openmp} or Intel TBB \citep{tbb}. OpenMP introduced in 1998 by Dagum and Menon \citep{dagum} is an industry standard since version 5.0 and is suported by most open source or commercial compilers. OpenMP is available in R itself if it is build with this option from the beginning, but it is dependent on the specific CPUs and C/C++ compiler. It can be also used in R by adding C++ processing functions through the Rcpp package \citep{rcpp1, rcpp2, rcpp3}. Intel TBB is a technology similar to OpenMP but is available only via C++. The RcppParallel package \citep{rcpppar} is a wrapper around Intel TBB library, making it easily accessible for R programmers. Both technologies allows users to build processing functions that make use of all the available cores of the processor on their desktop, speeding up the computations when large data sets are involved  or computationally intensive algorithms are used.\\

These technologies are somehow compiler dependent and not available for every user. To overcome this difficulty, now the base R  incorporates the \textbf{parallel} package that makes transparent for  users the low level operations to support shared memory parallelism. For example, \texttt{mclapply} function is the parallel version of the serial \texttt{lapply} and it applies a function to a series of elements, running them in parallel in separate processes, with the advantage of having all the variables from the main R session inherited by all child processes. However, the truly parallel execution  of the function on different data items is implemented only on systems that implements the FORK directive, i.e. Unix based systems. Windows doesn't support forking, thus \texttt{mclapply} and similar functions will be run in the sequential mode. Nevertheless, parallelization is still possible in this case too, using cluster processing, a model where a set of R processes run in parallel independently. Functions like \texttt{parLapply} or \texttt{parMapply} are using this model of execution to run processing functions in parallel but it poses on the user the task of sharing the variables among worker processes. Besides the \textbf{parallel} package there are other R packages that implement this kind of parallelism: \textbf{doMC} \citep{domc}, \textbf{doParallel} \citep{dop}, \textbf{foreach} \citep{frch}, \textbf{snowi} \citep{snow}.\\

The distributed memory parallelism uses a model called Message Passing which is described in the Message Passing Interface (MPI) standard \citep{mpi}. Widely used implementations of this standard include OpenMPI \citep{open_mpi} or MPICH \citep{mpich}. MPI involves a set of independent processes that run on their processor, directly accessing the data in that processor's memory. Communication between processes is achieved by means of sending and receiving messages. The communication operations between processes is the main bottleneck of this model, processing speed being usually much higher than sending or receiving data through the communication network. Less communication operations, the higher speedup will be obtained. This model has a main advantage over the shared memory model: it scales very well. Thousands of processors could be added to such a computer, obtaining really impressive computing power. Developing programs that use this paradigm involves writing them usually in C or Fortran and then linking them against an MPI library and then run them in a special configured environment. This is not an easy task to do for a statistician, but R packages like \textbf{Rmpi} \citep{rmpi}, \textbf{snow} \citep{snow} or \textbf{doMPI} \citep{dompi} present a high level interface to the user, hiding the complexity of message passing parallel programming.\\

As mentioned before, if we want to integrate new types of data sources into the statistical production the classical inferential paradigm has to be changed  and new methods involves using algorithms from machine learning or artificial intellingence area. A survey of the machine learning techniques currently used across different statistical offices can be found in \citep{mlos}. R packages like \textbf{rpart} \citep{rpart}, \textbf{caret} \citep{caret}, \textbf{randomForest} \citep{rforest}, \textbf{nnet} \citep{nnet}, \textbf{e1071} \citep{e1071} or Python  libraries like \textbf{Scipy} \citep{scipy},
\textbf{Scikit-learn} \citep{scikit-learn}, \textbf{Theano} \citep{theano}, \textbf{Keras} \citep{keras},  \textbf{PyTorch} \citep{pythorc} are among the tools that are best suited for the statistical production. Large frameworks like TensorFlow \citep{tensorflow} or Apache Spark \citep{apache} can also be used but they require specific skills from computer science area and have a steep learning curve, but connectors for R and Python are available that make those frameworks easier to use by statisticians.\\

Processing methods that make use of machine learning algorithms are frequently computing intensive. One solution to obtain reasonable running times even for large data sets is to use some parallel programming techniques and software packages already mentioned that exploit the multicore or many core feature of the commodity systems, but together with them another parallel computing paradigm called General-Purpose Computing on Graphics Processing Units (GPGPU) that was first experimented around 2000-2001  \citep{gpgpu} could be a viable solution. Today's GPUs have FLOP rates much higher than CPUs and this comes from the internal structure of a modern GPU: it has thousands of computing units that can operate in parallel on different data items, thus obtaining high throughputs. A detailed discussion of this computing model is far from the scope of current paper, but an interested reader can consult, for example, the work by \citet{gpgpu2}. CUDA \citep{cuda} or OpenCL \citep{opencl} are frameworks that allow users to build applications to take advantage of the immense computing power of the graphics processing units (GPU). Usually, they require applications written in C/C++ or Fortran and one may say that this is a task for a computer scientist not for a statistician, but in the last time several R or Python libraries have been developed to make the GPU accessible from these working environments familiar to statisticians. We can mention  \textbf{gmatrix} \citep{gm}, \textbf{gpuR} \citep{gpur1, gpur2}, \textbf{gputools} \citep{gputools} or \textbf{cudaBayesreg} \citep{cudabayes} R packages and \textbf{PyCUDA}  \textbf{PyOpenCL} \citep{pycuda},or \textbf{gnumpy} \citep{gnumpy} Python libraries that can be used to speedup the computations involved by different processing procedures.\\

Dedicated  systems are the other alternative when very large volumes of data need to be processed. One of the first dedicated computing systems tailored to make experiments with large data sets in official statistics was the UNECE Sandbox \citep{sandbox} which was a shared computing environment consisting in a cluster of 28 machines running a Linux operating system, connected through a dedicated high-speed network and accessible via a Web interface and SSH. This computing environment was created with support from the Central Statistics Office of Ireland and the Irish Centre for High-End Computing. Several large datasets where uploaded to this system from different areas: scanner data to compute price indices, mobile phone data for tourism statistics, smart meter data for computing statistics on electricity consumption, traffic loops data for transportation statistics, online job vacancies data and data collected from social media.  The software tools deployed in this environment were entirely new to the world of official statistics: Hadoop \citep{hadoop} for storing the data sets and performing some processings, Apache Spark for data analytics and Pentaho \citep{pentaho} for visual analytics. Together with them, the R software environment was also installed in the cluster.\\

Hadoop is a free software framework with the aim of storing and processing very large volumes of data using clusters of commodity hardware. Hadoop was developed in Java and thus, Java is the main programming language for this framework, but it can also be interfaced with other languages too, like R or Python. Although it is a freely available software, there are some commercial distributions that offer an easy way to install and configure the software as well as technical support. The most  widespread distributions are HortonWorks (that was used for the UNECE Sandbox) and Cloudera.\\

Hadoop framework includes a high performance distributed filesystem (HDFS - Hadoop Distributed File System), a job scheduling and cluster resource management component - YARN, and MapReduce which is a system for parallel processing of very large data sets. MapReduce implements a distributed model of computation that was first developed and used by Google \citep{mapred}.\\

Briefly speaking, Hadoop provides a reliable distributed storage by means of HDFS and an analysis framework implemented using the MapReduce engine. It is a highly scalable solution being able to run on a single computer as well as on clusters of thousands of computers. Large files are splited into blocks stored on different Data Nodes, while a Name Node is responsible with operations like opening, closing or renaming these files. MapReduce is a model of processing very large data sets on clusters of computers, first splliting the inputs in several chunks processed in parallel by the \textit{map} tasks. The results of the \textit{map} tasks are then forwarded to the \textit{reduce} tasks that perform an aggregation operation on them. All the complexity of the parallel execution of these tasks are hidden from the user that sees only a simple model of computation.\\

Hadoop framework was very successful for handling large data sets because of its high degree of scalability, flexibility and fault tolerance. It can be installed on commodity hardware or on supercomputers too, allowing massive parallel processing. It is able to store any kind of data, structured or not, and it is tolerant to hardware failures being able to send the tasks of a failed node to other live nodes. The files are stored in HDFS using a replication schema to ensure fault tolerance. Starting from the idea that is it  easier to move the computations than the data, when a computing node fails, the computations are send to another node that stores a replica of the data in the failing node.\\

\begin{figure}[h]
	\centering
	\includegraphics[width=8cm]{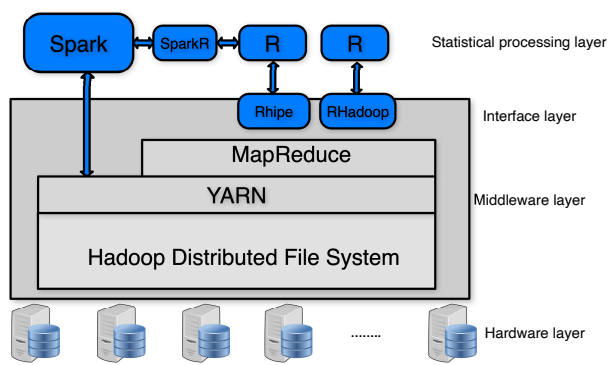}
	\caption{Hadoop infrastructure} \label{fig:hadoop}
\end{figure}

For statistical purposes, only Hadoop itself is rather difficult to be used, but when it is interfaced with usual statistical software like R, it becomes a powerful tool in the hand of statisticians. A typical architecture with Hadoop, Spark and other statistical tools is depicted in figure \ref{fig:hadoop}. Accessing the power of parallel processing of Hadoop from R is achieved through an interface layer made up from specialized R packages like \textbf{Rhipe} \citep{rhipe} or the collection gathered under the name of \textbf{RHadoop} \citep{rhadoop}.\\

Apache Spark is also an open source distributed computing framework, and it is newer then Hadoop. It provides a faster data analytics engine than the Hadoop MapReduce because it processes all the data in-memory. While Hadoop is better suited for batch processing, Spark also supports stream processing. It can be installed on a HDFS (like in figure \ref{fig:hadoop}) or as a standalone software. The \textbf{SparkR} \citep{sparkr} package provides a lightweight interface to use Spark from the R environment, making it easily accessible to statisticians. Spark has libraries that implement machine learning algorithms, graph analytics algorithms, stream processing and SQL querying. Spark and its very fast machine learning algorithms implementation proved to be a very useful tool especially for new data sources that require a model based approach.\\

\subsection{Push-out computation paradigm}
\label{sub:PushOutCompParad}

Statistical offices from the ESS started to implement their own in-house infrastructures to support processing needs for new data sources. We can mention here the ISTAT Big Data IT Infrastructure that consist in a 8-node Hadoop Cluster with Apache Spark as an analytics engine and Apache Impala for querying large amounts of data \citep{istat} or CBS Big Data Centre to name only two of them.\\

But soon after the initial enthusiasm of using new data sources,  the barrier of data access and the high costs stopped further in-house developments of IT infrastructures. Most of the new data sets are privately held data, and the data owners are reluctant when it comes to give access for statistical offices to their data. Moreover, the costs of such infrastructures are high and a single organization cannot support them on a long term. We assist in the last years to a paradigm shift: instead of developing huge IT infrastructures in-house,  using the cloud services available today at a lower cost seem to be a better solution. One of the first steps in this direction were made by European Commission with the Big Data Test Infrastructure  (\url{https://ec.europa.eu/cefdigital/wiki/display/CEFDIGITAL/Big+Data+Test+Infrastructure}) that was used in statistics for experimentation purposes during NTTS 2019 conference and after that for ESSnet Big Data project (\url{https://webgate.ec.europa.eu/fpfis/mwikis/essnetbigdata/index.php/ESSnet_Big_Data}). This infrastructure was built on Amazon Web Services cloud environment with a special configuration for statistical projects. It provided an Elastic Map Reduce (EMR) platform for big data processing built around the Hadoop ecosystem. Among the tools made available to the users on this platform we mention Apache Spark and Apache Flink for distributed data processing, Apache Hive and Apache Pig for querying the data, TensorFlow and Apache Mahout for machine learning applications, Apache Hue as a visual user interface, Jupyter Notebooks, R, RStudio, RShiny and an instance of MySQL relational database.\\

Another innovation that helps official statistics to overcome the data access barrier is \textit{to push the computations out} , at least partially \citep{fabio1}. Thus, instead of pulling in the data sets from private companies and process them using in-house computing systems, the data are not moved from the data holders' premises, but stay there and are only used by official statisticians, running commonly agreed algorithms on private companies' computing systems and getting only some form of aggregated results. This will avoid sharing the privately held microdata which most of the time are an invaluable asset for companies and raises complicated legal problems. However, there are concerns that official statisticians are not in control of the processing stage and results may be biased or the quality of the results would not be as expected.\\

\begin{figure}[h]
	\centering
	\includegraphics[width=8cm]{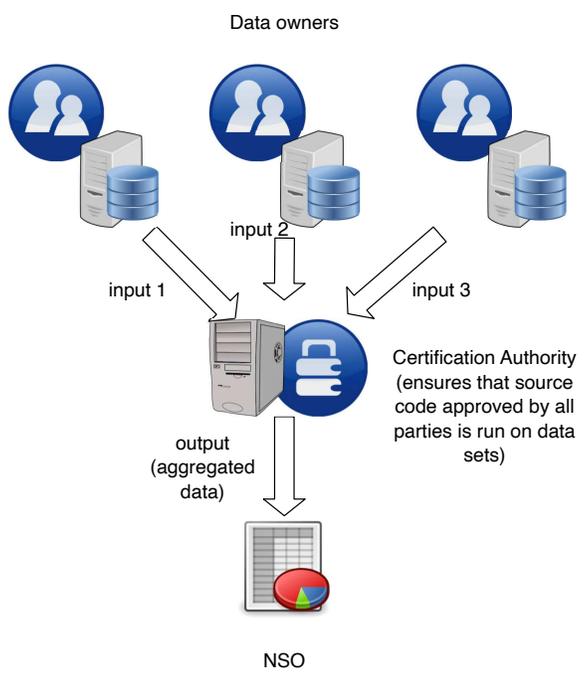}
	\caption{Trusted computation} \label{fig:ratss}
\end{figure}

To overcome this problem, a sort of certification authority trusted by all parties could be involved and thus, the processing algorithms would be transparent and trusted by all parties. This is one of the ideas on which the Reference Architecture for Trusted Smart Statistics proposed by \citet{fabio1} is build upon. In figure \ref{fig:ratss} we show this idea:  several data owners and the statistical office agree upon the algorithms for data processing and the Certification Authority is a guarantee that only the agreed source code is run on the data sets.\\

\begin{figure}[h]
	\centering
	\includegraphics[width=8cm]{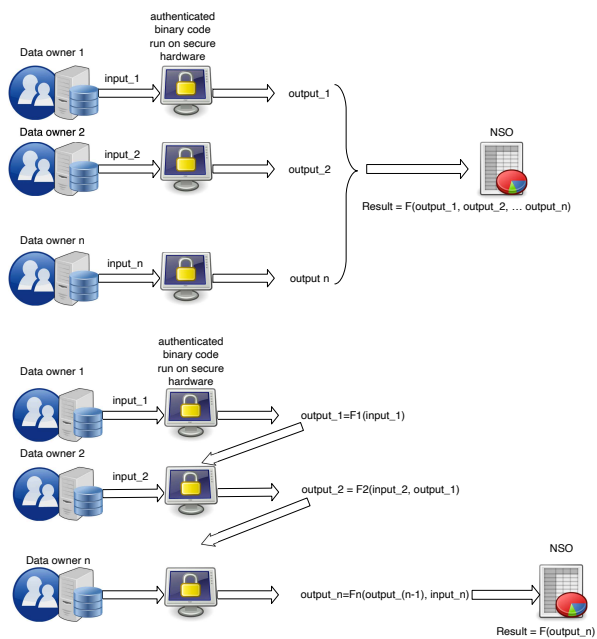}
	\caption{Trusted computations with multiple data owners} \label{fig:se}
\end{figure}

The simplest case is when only one data owner is involved in such a process. In this case running an authenticated binary code in a secure (trusted) hardware environment could solve the problem of ensuring that indeed the code that was executed on the data sets is exactly the code that was agreed between the parties (in our case the data owner and statistical office). This model can be generalized when more data holders participate in this process and the final result can be obtained either by taking the partial outputs of an agreed code (function) applied on each data set separately and then composing them using another function by the statistical office, or by chaining the partial results using again the agreed functions implemented in an authenticated binary code and run in a secure hardware environment \citep{fabio2}. These two cases are presented in figure \ref{fig:se}. In the upper part of the figure we have the case when each data owner provides a data set $input_i$ that is processed by an agreed function in a secure hardware environment and the results $output_i$ are then fed into a function $F$ that computes some aggregated measure.  In the bottom part of the figure we have the case when the output of the first processing algorithm is sent as an input to the second algorithm and so on. Again, an authenticated binary code and a secure hardware environment provides all that is neccesary to be sure that the code executed on data sets is the one that was agreed between the data owners and statistical office.\\

The technologies needed for such a mechanism are known and widespread. Code signing is a form of binary authentication that can be used in this case and the Trusted Execution Environment (TEE) standard \citep{tee} could be considered a potential candidate, all major hardware producers (Intel, AMD, ARM) providing support for TEE implementations \citep{inteltee, amdtee, armtee}. Mainly, all modern processors provide a mechanism that allows a process to be run in such a way that its data is not seen by other processes or even by the operating system.\\

The case when the final statistical aggregates / estimates suppose to combine data sets from different owners and these combined data are then sent as input to a function that computes the estimates requires more elaborated processing techniques borrowed from the Privacy-Preserving Computation Techniques (PPCT), a hot research field that combines classical cryptography with distributed computing, to provide protection for data owners and in the mean time allowing statistical analyses to be performed \citep{unppct}. Using such techniques allows one to perform data analyses on data sets coming from different owners while the data remains opaque to all the parties involved, thus obtaining an end-to-end protection of the data. Nevertheless, these techniques have their own implementation costs, regarding both the hardware and software investments, that cannot be neglected.\\

One of the proposed PPCT to be used by statistical offices in cooperation with data owners is the Secure Multi Party Computation (SMPC) \citep{fabio3, unppct}. SMPC is about jointly evaluating a function that all parties agreed upon using a set of different inputs coming from several parties and maintaining the confidentiality of the data so that no participant can have access to the raw data provided by the others. This technique divides the input data into random shares that gives back the original data if they are combined, and these shares are then distributed among all the participants. The shares can then be combined to produce the desired output.\\

Formally, SMPC deals with a set of participants, $p_1, p_2, \dots p_n$, each of them having a data set $input_1, input_2, \dots,  input_n$, who intend to compute a function $F(input_1, input_2, \dots,  input_n)$ keeping the inputs secret.  An SMPC protocol will ensure all participants about the input privacy (i.e no information can be inferred by a party about others party' data)  and correctness of the output. While the first attempts to develop such a computation protocol dates from early 1982 when a secure two-party protocol has been introduced by \citet{yao1} and then further developed and formulated in 1986  \citep{yao2} SMPC is still an academic research topic nowadays and commercial solutions for this protocol are still in an early stage. Nevertheless, as information technology is advancing with a very fast speed, this could be a viable solution official statistics.\\

\begin{figure}[h]
	\centering
	\includegraphics[width=8cm]{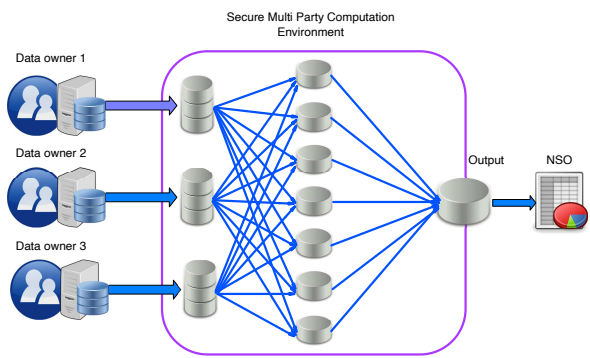}
	\caption{Secure Multi Party Computation} \label{fig:smpc}
\end{figure}

In figure \ref{fig:smpc} we showed a schematic example of this technique where several data owners provide their data to an SMPC environment where they are processed and an output is provided to an statistical office.\\

Other privacy-preserving computation techniques proposed to be used in official statistics are the Homomorphic Encryption, Differential Privacy or Zero Knowledge Proofs \citep{unppct}. However, all these techniques requires further experimentation and development of practical software implementations.

\section{Skills, human resources, management}
\label{sec:SkillHR}
This section is remarkably opinative and provocative on purpose to raise thought and debate. Certainly, the analysis above does not pretend to be exhaustive and can be further completed with deeper and more extensive reflections on some mentioned items or new ones. In any case, the success in the adoption of new solutions and changes in the statistical production necessarily requires new skills and an extraordinary exercise of management.\\

To begin with, at odds with common belief, we claim that the production of official statistics in a statistical office is an activity closer to Engineering than to Social Science and Statistics. By no means this signifies that Social Science and Statistics are marginally needed. Experts on National Accounts, on Demography, on sampling, etc.\ are absolutely necessary, but in the same way as being an expert physicist in the electromagnetic field and the law of induction does not make you capable of producing and distributing electrical power to every dwelling in a country, the knowledge in those disciplines does not guarantee the industrial production of official statistics comprising a National Statistical Plan. This need for an engineering view of official statistical production to cope with complexity was already made patent with the advent of international production standards at the beginning of the 21st century. We are convinced that with new data sources, especially digital data, this approach is urgently required.\\

Consequently, a new organization of the production processes brings new skills into scene. Some traditional skills will need to be superseded and some others reformulated or adapted to the new production conditions. However, we view this as an integration process not as a general disruptive substitution of techniques, procedures and routines, in general. The use of information technologies and computer science needs to permeate the production and sometimes this may produce a cultural resistance to change (\textquotedblleft statisticians do not have to program computer systems because that task belongs to another academic discipline\textquotedblright, say some in private). Should archaeologists avoid incorporating knowledge about carbon dating and DNA analysis into their work because these belong to other disciplines? They may not need how to conduct themselves a DNA analysis managing a DNA sequencer, but certainly their renewed skills allow them to openly communicate with DNA experts and modernise their work accordingly.\\

When all these new skills are mentioned in future prospects of Official Statistics, the focus is instinctively placed on technical or junior staff, possibly thinking of new recruitment and plans of continuous training. This is obviously an element to be considered but we find it more critical the extension of these skills and a clear understanding of their consequences in production among the top management of the organization. If they need to take critical decisions, they also need to clearly understand some technical and organizational details about the implications of these decisions. For example, moving away from a stove-pipe production model inefficiently divided into silos to a standardised production model sharing methods, tools, data architecture, process design, etc.\ necessarily brings changes into the chart of the organization and the governance structure. How does it all smoothly fit to work in practice? These are difficult questions rooted in some technical aspects with consequences throughout the whole organization.\\

Furthermore, in many statistical offices, there are scarce resources fully devoted to production in a highly demanding environment with little room to acquire these new skills. In many cases, the computer science, ICT, and programming background is even outdated (for these same reasons). The training modernization plans, in our view, should consider this staff also as a primary target. For example, the introduction of new distributed computing systems with object-oriented and functional programming languages is clearly necessary, but it is also that necessary to bring senior staff to the point in which these training programmes are also accessible and valuable for them. With this new knowledge, they can provide highly valuable insights into the modernization process.\\

In this line, newly recruited staff should be demanded to fulfill this joint profile with both computer science and statistics skills. Interestingly enough, as in other industries (e.g. finance), a lot of value can be gained from professionals with different backgrounds such as engineers, physicists, chemists, \dots because of their system modelling abilities. In any case, professional training needs to be continuous and embracing cross-cuttingly all the staff, since technologies are changing very fast now.\\

Management challenges do not end with human resources and new skills. With traditional survey data, the complete production process fall to statistical offices from survey design over data collection to production and dissemination. With administrative sources, data are already generated independently of the statistical purposes and specific agreements with other public bodies must be settled to access and use them to produce official statistics. With digital data in private hands, the new scenario portrays a higher entangled situation. Data holders in the private sector will be necessarily part of the statistical production process and this entails an extraordinary exercise of management on data, quality, metadata, trust, technology, \dots\\

Furthermore, in a datafied society with an increasing economic sector based on data, information, and knowledge, statistical offices need to decide which role to play in an environment with multiple actors, which turn out to be both data holders and stakeholders of a generalised statistical production. Statistical offices will never be the unique producers of statistical outputs with social interest. Which relation to these products are statistical offices to take on? Options do exist. Statistical quality certification to offer quality assurance is a possibility. The enrichment of data and/or methodologies in the private production processes can also be considered. In any case, all these options entail new exercises of management and leadership.

\section{Conclusions}
\label{sec:Con}

Data sources for the production of official statistics can be grouped in survey data, administrative data, and digital data. The advent of both administrative and digital data introduces important changes in the production landscape of statistical offices. The lack of statistical metadata (data are generated prior to any statistical purposes consideration), the economic value of data, and their ownership by third people and not by data holders characterise these new data sources. These have implications for data access/use, for statistical methodology, for quality, for the IT environment, and for management. For every aspect several issues need to be considered. As summary statements we can conclude the following:

\begin{itemize}
    \item In our view, public-private partnerships stand as the preferred option to incorporate new data sources into the routinely production of official statistics. These partnerships must consider aspects from all perspectives. Guarantees for privacy and confidentiality must be pursued at all costs. Official Statistics already has a tradition in this line since design-based inference needs unit identifiability and statistical disclosure control techniques are increasingly more sophisticated. In our opinion, legislative initiatives to provide legal support, if further needed, must be undertaken from this partnership point of view. New disciplines as cryptology need to be introduced.
    \item Sampling designs cannot be useful to face the inferential step with the new data sources. Traditional survey methodology, however, should be seen as an inspiration to pursue accurate estimators. The notion of sample representativeness, not being a mathematical concept, is still to be understood as the search for estimators with low mean square errors (or similar figures of merit), as survey methodology actually does. Probability theory is still the best option to deal with inference.
    \item Machine learning and artificial intelligence seems of limited use for the inferential stage, since we never know the ground truth for the learning step training algorithms. However, this is not the case for multiple production tasks along the production process. Indeed, the wealth of traditional survey data and paradata stands as an opportunity to make use of these techniques in the production process, especially regarding the lack of statistical metadata in the new data sources.
    \item Current quality frameworks are strongly survey-oriented. Although quality dimensions in Official Statistics appear still to be valid, the subtleties arising from the new nature of data need to be considered both in their definitions and in the indicators derived thereof.
    \item Special focus should be placed on relevance. New insights can be a priori gained from the wealth of new data (e.g.\ investigating the interaction between population units). Thus, new statistical outputs must be devised.
    \item New hardware and software environments are needed to incorporate new data sources into the production. Open source software ecosystems like R or Python together with the accompanying libraries for official statistics seem to be the future of the statistical data processing. The hardware infrastructures are changing too. While few years ago several statistical offices built their own (in-house) computing systems they proved to be very costly and now we are witnessing a new trend, i.e. usage of cloud-based hardware infrastructures. These systems are equipped usually with specific big data software products like Hadoop or Apache Spark. However, in the IT field technologies are changing with an unprecedented speed and is difficult to predict which technology is the best for statistical purposes.
    \item A crucial challenge to cope with the implications brought by new data sources is the integration of all the preceding facets into a renewed production system. This demands an extraordinary exercise of management and leadership. Statistical offices, in our view, should strive to assume a leading role in the new datafied society.
\end{itemize}

\section*{Acknowledgments}
The views expressed in this paper are those of the authors and do not necessarily reflect the views of their affiliating institutions.

\bibliographystyle{chicago}

\end{document}